\documentclass[prl,amsmath,amssymb,reprintnumbers,nofootinbib, twocolumn]{revtex4}
\usepackage{latexsym}
\usepackage{epsfig}
\usepackage{amssymb}
\usepackage{amsmath}
\usepackage{color}
\usepackage{subfigure}
\usepackage{comment}

\newcommand{\be}{\begin{eqnarray}}
\newcommand{\ee}{\end{eqnarray}}

\begin{document}

\preprint{}
\title{Reply to C. Tsallis' Comment on our ``Nonadditive Entropies Yield Probability Distributions with Biases not Warranted by the Data"}
\author{Steve Press\'e$^{1}$, Kingshuk Ghosh$^{2}$, Julian Lee$^{3}$, Ken A. Dill$^{4}$}
\date{\today}
\affiliation{ 
$^1$ Department of Physics, Indiana Univ.-Purdue Univ. Indianapolis, Indianapolis, IN 46202. \\
$^2$ Department of Physics and Astronomy, University of Denver, CO 80208.\\
$^3$ Department of Bioinformatics, Soongsil University, Seoul 156-743, Korea.\\
$^4$ Laufer Center for Physical and Quantitative Biology and Departments of Physics and Chemistry, Stony
Brook University, NY 11794. }
\maketitle


 In Ref.~\cite{presseprl}, we invoked the powerful results of Shore and Johnson (SJ) \cite{shore}, who showed that, under quite general circumstances, the least-biased way to infer a probability distribution $\{p_i\}$ is to maximize the Boltzmann-Gibbs relative entropy
\be 
H = -\sum_{i}p_{i}\log(p_{i}/q_i)
\label{eq1}
\ee
(or any function that is monotonic with this entropy), under constraints where $q_i$ is the prior distribution that contains any foreknowledge of the system.  In Ref.~\cite{presseprl}, we showed that mathematical forms of 
$H$ that are not monotonic with Eq.~(\ref{eq1}) -- which we call noncanonical entropies -- lead to unwarranted biases.  In his comment to our Ref.~\cite{presseprl}, Tsallis objects to this argument on three grounds.  We show here the flaws in his objections.

First, Tsallis contends that ``nonadditive entropies emerge from strong correlations which are out of the SJ hypothesis" adding that the ``SJ axiomatic framework addresses essentially systems for which it is possible to provide information about their subsystems without providing information about the interactions between the subsystems, which clearly is {\it not} the case where nonadditive entropies are to be used".  These statements are incorrect.  SJ is not limited in any way to small interactions or weak correlations; it can handle interactions of any strength.  

Standard methods of statistical physics -- which are grounded in Eq.~(\ref{eq1})  -- 
provide a clear recipe for treating correlated systems while assuming no initial subsystem correlations. For example, we don't need a special form of Eq.~(\ref{eq1}) in order to build spin-spin correlations into an Ising model. On the contrary, the Boltzmann weights for a spin-spin-correlated Ising model
are constructed by assuming that spin correlations originate from the data (which is used
to constrain Eq.~(\ref{eq1})). In the absence of spin-spin correlations,  
standard statistical physics returns an Ising model with decoupled spins, as it should.
On the other hand, if we choose to assume {\it a priori} some particular coupling between the spins that do not originate from data, 
then SJ prove that it should be introduced through $\{q_i\}$ exactly as it appears in Eq.~(\ref{eq1}).  In particular, models of power laws can only arise in a principled way either from data constraints or from $\{q_i\}$ in conjunction with Eq.~(\ref{eq1}).  There is no principled basis for power laws that can be obtained by re-assigning the meaning of $H$ and changing its form \cite{presseprl}.

 Second, Tsallis asserts that ``the SJ axioms, demanding as they do, system independence, ...''.  This is not correct.  The SJ inference procedure is not limited to independent systems.  Rather, SJ asserts that \emph{when} systems \emph{are} independent of each other, \emph{then} the joint outcome for two independent systems must be the
product of marginal probabilities if data are provided for systems independently \cite{rmp}.  SJ is otherwise perfectly applicable broadly across situations not involving independent systems.
 
 Third, Tsallis asserts that ``This is the deep reason why, in the presence of strong correlations, the BG entropy is generically not physically appropriate anymore since it violates the thermodynamical requirement of entropic extensivity.''  In this statement, Tsallis is incorrectly asserting that extensivity must be a foundational property from which the functional form of the entropy follows.  In fact, the logic is just the opposite.  Extensivity -- or not -- of an entropy is the outcome of an inference problem at hand, not its input.  Certainly throughout much of equilibrium thermodynamics, extensivity happens to hold.  But that is a matter of the particulars of that particular class of applications. 
 
 Said differently, this argument confuses the distinction between \emph{pre-maximization} and \emph{post-maximization} entropies.  SJ focus on pre-maximization.  SJ seek a functional $H$ that, upon maximization, achieves certain properties required for drawing consistent unbiased inferences.  At this stage, no system property (such as how entropy scales with system size) is yet relevant. This is just establishing a very general inference principle.  However, once maximization has been performed, $H(\{p_i = p_i^*\}) = -\sum_i p_i^* \ln p_i^*$ is an entropy function that may be used to make predictions about physical systems including how properties scale with system size.  The SJ argument is agnostic about whether extensivity holds or not for the post-maximization entropy 
 $H(\{p_i = p_i^*\})$.
   
  In short, the great power of the SJ arguments is in showing that Eq.~(\ref{eq1}) is an extremely broad and deep result, applicable across all matters of inference of probability distributions, given only priors ($q_i$) and given new information.  The power of the SJ arguments is that they apply upstream of any particular application, whether it should involve equilibria or dynamics, materials or informational channel capacities or other, weak or strong correlations, extensivity or not, or any other particularity.  We are assured by Ref. \cite{shore} that no other form of entropy function -- beyond those monotonic with $H$ -- can generate unbiased inferences.

\hspace{-0.15in}{\bf Acknowledgements:} 
SP acknowledges support from the NSF (MCB Award No. 1412259).
KD acknowledges the support of NIH grant 5R01GM090205-02 and the Laufer Center. 
KG acknowledges support from the Research Corporation for Science
Advancement as a Cottrell Scholar.


\bibliographystyle{unsrt}

\end{document}